\documentclass[final,3p,times,twocolumn]{elsarticle}

\usepackage{graphicx,stfloats}
\usepackage{color}

\usepackage{amstext,amsmath,amsfonts,amssymb}
\usepackage{epstopdf}
\newcommand{\norm}[1]{\left\lVert#1\right\rVert}

\biboptions{sort&compress}
\usepackage{natbib}

\journal{Proceedings of the Combustion Institute}

\begin{document}

\begin{frontmatter}

\title{Ensemble-LES Analysis of Perturbation Response of Turbulent Partially-Premixed Flames}

\author[label1]{Malik Hassanaly\corref{cor1}}
\ead{malik.hassanaly@gmail.com}

\author[label1]{Venkat Raman}

\address[label1]{Dept. of Aerospace Engineering, University of Michigan, Ann Arbor, MI-48109}
\cortext[cor1]{Corresponding author:}

\begin{abstract}
The response to small perturbations of the Sandia CH$_4$/air flame series is studied using a dynamical systems formulation. Here, the Sandia D and E turbulent partially-premixed flames are computed using large eddy simulation (LES) with a flamelet/progress variable approach. Using 300 simultaneous LES computations of each flame, the partial Lyapunov spectrum of the flow is obtained, which provides a set of Lyapunov exponents (LEs) and Lyapunov vectors (LVs). Special numerical procedures to handle such large data sets and perform matrix manipulations on-the-fly are discussed. From the LEs, the dimension of the attractor for the flames was estimated, for the first time, to be at least 5000 (flame D) and 10500 (flame E). This is a significant result indicating that the chaoticity of the turbulent flame is confined to a volume of phase-space that is quantified. The growth of perturbations occurs at time-scales that are much larger than the smallest strain-rate based times scales, but smaller than the integral scales. Further, the Lyapunov vectors that determine the perturbation response show that the strongest perturbations are aligned with the jet breakdown region, while other LVs are aligned with the shear layer. It is found that both the extinction/re-ignition process and the turbulence-induced jet motion strongly contribute to the flame chaoticity but are not equally important for all the LVs. These results provide new insights into the perturbation response of turbulent flames. More broadly, this Lyapunov approach allows a full characterization of the chaotic dynamics associated with turbulent flames, which have applications for both modeling and control of such complex flows.
\end{abstract}

\begin{keyword}

Lyapunov spectrum \sep Turbulent partially-premixed flame \sep Attractor \sep Dynamical systems \sep Rare events 
\end{keyword}

\end{frontmatter}
\clearpage
\setcounter{page}{1}

\section{Introduction}
The issue of flame stability is central to the design of practical combustion devices. Flame instabilities can occur due to multiple causes, such as thermoacoustic interactions, large scale flow variations such as vortex breakdowns, or increase in local strain and dissipation rates that can cause local and ultimately, global extinction. Since these flows are generally turbulent, the underlying physical processes are multi-scale in nature, and are often described statistically \cite{popebook}. An alternative view of understanding instabilities is by probing the response of the combustor to small perturbations. These techniques have been used for example to construct system identification tools \cite{noiray2017method,polifke2014black} and flame transfer functions \cite{palies2010combined}. Derived from dynamical systems theory, such tools are useful for isolating key dynamical modes, especially when the instabilities are driven by large scale energetic motions, or possess characteristic features that have frequencies much lower than those associated with the inertial and dissipative range of the turbulent energy spectrum. Interestingly, these tools are not geared towards isolating events that occur with very low frequency that can push the combustor towards an instability, often referred to as rare events. However, the use of perturbations to characterize the stability of a flame is a powerful concept that needs to be explored further. In particular, the focus of this work is to determine the response of practical turbulent flames to finite perturbations.

As a thought process, consider a turbulent jet flame. The complete perturbation response will involve making small changes to each variable at each location in the flow and measuring the response of the flame over a finite time. This is clearly intractable unless mathematically rigorous reduction approaches are used to reduce the number of such perturbations that are needed. In this context, dynamical systems theory provides a powerful framework. Here, the starting point is to treat the unsteady spatially-discretized governing equations for a turbulent flame as a set of coupled ordinary differential equations, with the number of variables $N = N_s \times N_g$, where $N_g$ is the number of grid points in the domain, while $N_s$ is the number of variables solved at each grid point. Hence, the time evolution of the system can be considered as a trajectory in the phase space spanning $N$ dimensions. Several concepts from dynamical systems theory have been applied to combustion systems in the past \cite{pope2010self,malikmcs10,malikaiaa2017,kaiser_crom}. More recently, these approaches have gained renewed interest due to breakthroughs in algorithms \cite{wolfe,ginelli} and the vast increase in computational power. Of particular interest is the construction of so-called Lyapunov vectors and exponents \cite{benettin2} that provide the fastest growing perturbations to the system. The Lyapunov exponents (LEs) provide the growth/decay rate of perturbations applied in the directions specified by the Lyapunov vectors (LVs). Conceptually, this is similar to Fourier decomposition, proper orthogonal decomposition (POD) \cite{berkooz1993proper} or dynamic mode decomposition (DMD) \cite{motheau2014mixed}, which constructs the original flow field in terms of basis vectors and coefficients. Unlike these techniques that approximate the flow field directly, however, the LV approach represents the response to perturbations to the flow field.

The LEs and LVs are related to the concept of a strange attractor \cite{ruelle} for chaotic systems. The long-time trajectories in phase-space for ergodic flows (i.e., statistically stationary flows) is confined to a volume that is termed the attractor \cite{temam,pope2010self}. For chaotic systems, such as turbulent flames, this attractor spans a smaller number of dimensions $N_d$, which is smaller than $N$. The LEs can be either positive or negative, implying that perturbations along the corresponding LVs either grow or decay, respectively. However, the number of positive LEs is directly related to $N_d$. In other words, only perturbations along a finite number of directions can lead to subsequent growth in those perturbations. Although $N_d << N$, it has been theoretically shown to scale with Reynolds number as $N_d \propto \text{Re}^\alpha$, where $\alpha \sim 2.25-3$  depending on the shape of the turbulence spectrum \cite{temam}. Hence, estimating the dimension of attractors for turbulent flames will provide tremendous insight into the dynamics of such flows.

In the past, LEs have been determined for a number of non-reacting turbulent flows \cite{keefe,mohan2017scaling}. Keefe et al.~\cite{keefe} determined the number of positive LEs for a turbulent channel flow. Mohan et al.~\cite{mohan2017scaling} determined the largest LE for homogeneous isotropic turbulence, and demonstrated that the inverse of LE (which provides a time-scale) decreases faster than Kolmogorov time scale. More broadly, the Lyapunov perturbation theory and related concepts have been widely used in the field of medium range weather prediction \cite{kalnayBook,kalnay,buizza}, where similar vectors are used to determine the uncertainty in predictions. In most applications, the focus is on the largest LE and the corresponding LV, which is computationally only twice as expensive as conducting the baseline simulation. However, to understand the dynamics of the turbulent flame, it is important to determine the series of exponents and vectors. Such computations are inherently expensive and are also subject to numerical errors \cite{keefe}, but recent advances have increased the reliability and allowed parallel computations on high-performance machines \cite{fernandez,maliknumericsLE}.

With this context in mind, the objective of this work is twofold: 1) to use Lyapunov theory to estimate the dimension of the attractor for the Sandia flames D and E, and 2) to analyze the instability modes provided by the Lyapunov spectrum for these partially-premixed flames with local extinction/re-ignition phenomena. The rest of the paper in organized as follows: In Sec.~\ref{sec:targetConf}, the target configuration is presented. The necessary background about Lyapunov theory is also described along with the numerical algorithm used to compute the Lyapunov spectrum. Section~\ref{sec:results} is devoted to the results of the Lyapunov analysis and their interpretation for the target configuration.

\section{Simulation configuration and Lyapunov computations}
\label{sec:targetConf}
In this section, the computational approach for obtaining a set of LEs and LVs is described. This technique requires a baseline simulation approach for describing the turbulent flame, and a multi-simulation algorithm to obtain the LEs and LVs. The target flame is the Sandia partially-premixed flame series \cite{barlow}, which has been extensively studied using a number of simulation approaches \cite{ramanpitsch-sandia1,pitschD,tnf4}.

\subsection{LES/flamelet approach for Sandia D/E Flames}

\begin{figure}
\center
\includegraphics[width=0.49\textwidth,trim={0cm 0cm 0cm 0cm},clip]{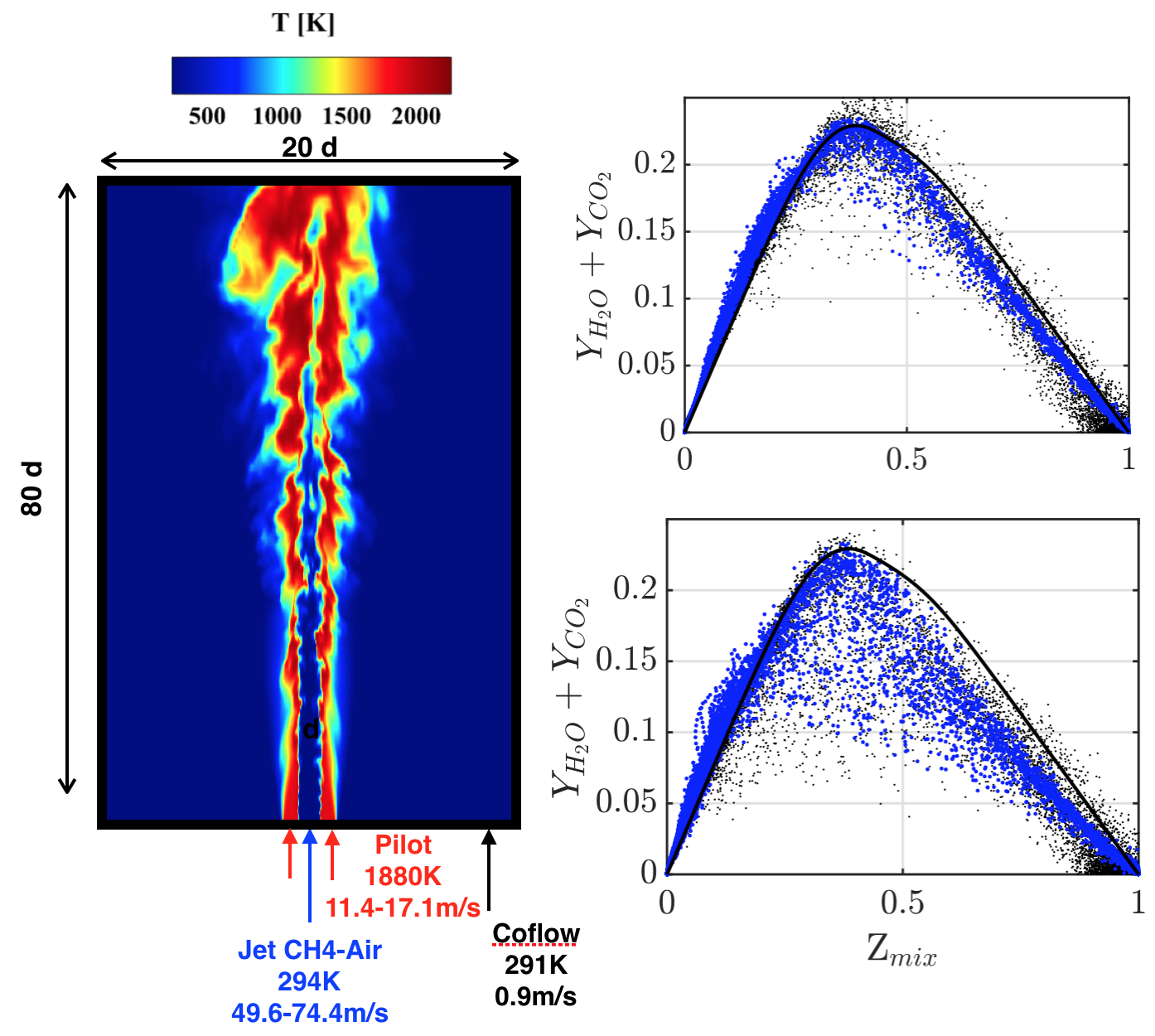}
\caption{Left: schematic of the simulated configuration. The contour does not show the entire domain. Right: scatter data of progress variable conditioned on the mixture fraction for the simulation (blue dots) and the experiments (black dots), for the D flame (top) and the E flame (bottom) at $x/d=7.5$. The plots are overlayed with a burning flamelet obtained at a strain rate $a = 87 s^{-1}$.}
\label{fig:sandiaIllustration}
\end{figure}

The Sandia flames use partially-premixed CH$_4$/Air mixture as fuel, and air as oxidizer. Figure~\ref{fig:sandiaIllustration} shows a schematic of the piloted flow configuration. The different flames in the series contain different inflow velocities for the central jet and the pilot, which change the local strain rates and the level of extinction. In this study, the D and E configurations with central jet bulk velocities of $49.6$ m/s and $74.4$ m/s and pilot bulk velocity of $11.4$ m/s and $17.1$ m/s, respectively, are considered. The D flame exhibits very limited extinction, while flame E exhibits significant local extinction phenomena especially in the near field at $x/d < 20$, where $d$ is the fuel jet diameter. 

The LES approach has been widely used to simulate these configurations   \cite{ramanpitsch-sandia1,pitschD} and will be utilized here. Although many different combustion models have been applied to this flame, the flamelet/progress variable approach (FPVA) \cite{pierce2004progress} will be used. The progress variable is defined as sum of mass fraction of water and carbon dioxide, $Y_{H_2O} + Y_{CO_2}$. The flamelet solutions are tabulated using 1D-counterflow diffusion flames computed with FlameMaster \cite{pitsch-flamemaster} using the GRI-2.11 chemical mechanism \cite{gri211}. The LES computations use a low-Mach number structured code with a staggered arrangement of variables \cite{desjardins2008high}. Transport equations for velocity, mixture fraction and progress variable are solved, and the mixture fraction variance is obtained using an algebraic closure model with dynamic coefficient \cite{kaul2013large}. Similar to Ref.~\cite{ramanpitsch-sandia1}, the cylindrical computational grid spans $80d$ in the axial direction ($256$ points) and $20d$ in the radial direction ($128$ points). In the azimuthal direction $32$ grid points are used. An additional simulation with a coarser grid of size $172 \times 90 \times 32$ was used to determine the impact of spatial resolution on the results.

Figure~\ref{fig:sandiaIllustration} shows key results that demonstrate the capability of the solver to capture local extinction. The conditional scatter plots show the presence of low temperature zones near stoichiometric conditions for the E flame, indicating local extinction events. The contour plot shows that even when such local extinction is present, there are re-ignition events downstream due to mixing with hot fluid pockets that prevent global extinction. Similar to the other LES/FPVA results, these simulations capture unconditional statistics accurately (not shown here). This simulation will form the baseline for the Lyapunov calculations described below.

\subsection{Computation of Lyapunov Exponents and Vectors}
\label{sec:lyapunovAnalysis}

\begin{figure}[h]
\center
\includegraphics[width=0.495\textwidth,trim={0cm 0cm 0cm 0cm},clip]{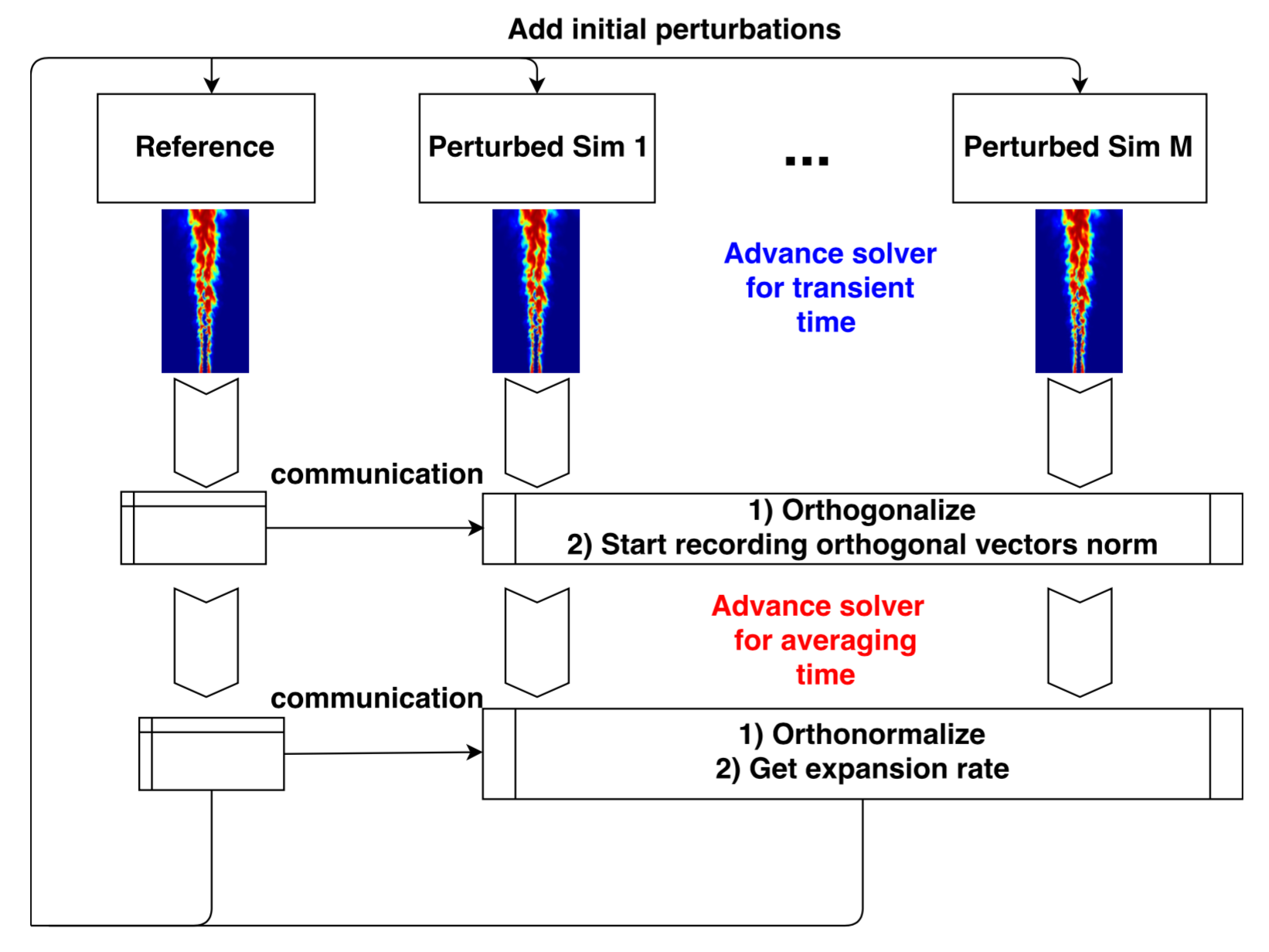}
\caption{Flow chart of the modified algorithm used to compute the Lyapunov spectrum of the D and E flames.}
\label{fig:flow}
\end{figure}

The dynamical system that describes the modeled governing equations for the flames is of dimension $N = 7.34 \times 10^6$, with $N_s = 7$ and $N_g = 256 \times 128 \times 32$. There exists then $N$ sets of LEs and LVs, but only a fraction of positive LEs that are indicators of the chaotic structure of the flow. The algorithm used here seeks the $N_l$ largest LEs and the corresponding LVs. $N_l$ is set by the user based on the availability of the computational resources, and the number required to reliably estimate the dimension of the attractor (to be discussed below). The numerical approach used here is based on the algorithm of Benettin et al.~\cite{benettin2}. The LVs that are computed are specifically called Gram-Schmidt vectors (GSVs), which form an orthonormal basis (incomplete if $N_l < N$) for any perturbation applied to the system. Where LVs were computed in past studies, only the first vector was determined \cite{vastanoMoser}. The algorithm used here is schematically shown in Fig.~\ref{fig:flow}, and involves simultaneous evolution of a number of LES computations, which interact at regular intervals to alter individual fields. It should be noted that the number of variables that is included in the state-vector to define $N$ is dependent on the LES solution algorithm. The low-Mach number solver used here employs a pressure-projection step to ensure mass conservation \cite{desjardins2008high}. Consequently, the pressure field is taken to be a function of the velocity field and not included in the state vector. For the combustion model employed here, the state vector is defined as $\phi = \{ \rho \boldsymbol{U}, Z_{mix}, C, Z'^2, \Delta \rho \}$, where $\rho \boldsymbol{U}$ is the momentum vector, $Z_{mix}$ denotes the mixture fraction and $C$ is the progress variable. $Z'^2$ is the mixture fraction variance and $\Delta \rho$ is the density variation at each time-step. Perturbations are applied at each interior point but at the boundary. This is necessary to preserve the structure of the attractor. A more detailed description of the choice of variables is provided in Ref.~\cite{maliknumericsLE}. 
The numerical method adopted here for the computation of the LE is applicable to absolutely unstable system which is in general not guaranteed for open systems like jet flames. However, several studies in the past have indicated that jets with large density ratio are more subject to absolute instabilities than convective instabilities \cite{huerre1990local,sreenivasan1989absolute}. To ensure that the jet studied here is absolutely unstable, perturbations were applied only near the outflow. It was found that even these perturbations were transported upstream due to the low-Mach number nature of the solver, and ultimately grew in magnitude. This conclusively showed that the Sandia jet flames exhibit asbolute instability.

The algorithm is implemented as follows:
\begin{itemize}
    \item \underline{Step 1:} Start from a statistically stationary solution of the flow denoted by $\phi_0$, where $\phi_0 \in {\mathbb{R}^{N \times 1}}$; Choose $N_l$ random not aligned vectors $\delta \phi_i \in \mathbb{R}^{N \times 1}, i =1,\cdots,N_l$. Each random vector is the first perturbation applied to each simulation. At future timesteps, the applied perturbations are forced to be aligned with the computed GSV.
    \item \underline{Step 2a:} Start iteration loop $k$. Initialize $N_l$ simulations with initial conditions $\phi_i = \phi_0 + \epsilon \delta \phi_i, i = 1,\cdots,N_l$, where $\epsilon$ is a perturbation norm that is a user-defined parameter that is set to $10^{-5}$ here. The impact of the choice of $\epsilon$ on GSV computation has been studied elsewhere \cite{malikaiaa2017, maliknumericsLE}, with the above choice considered optimal for the current configuration.
    \item \underline{Step 2b:} Evolve $N_l$ perturbed simulations and the baseline simulation simultaneously for time $t = \tau_1$  to obtain solutions $\phi_i^{\tau_1}$ and $\phi_0^{\tau_1}$; Compute $\delta \phi_i^{\tau_1} = \phi_i^{\tau_1} -\phi_0^{\tau_1}, i = 1,\cdots,N_l $
    \item \underline{Step 2c:} Orthogonalize the perturbation vectors $\delta \phi_i^{\tau_1}$ using the classical Gram-Schmidt procedure repeated until convergence \cite{hernandez2005orthogonalization}. The purpose of this first orthogonalization is to accurately record the growth of orthogonal subspaces after allowing sufficient time for the trajectories to reach the attractor \cite{maliknumericsLE}. Record the initial norm of the orthogonalized perturbations $ \norm{\delta \phi_i^{\tau_1}}, i = 1,\cdots,N_l$.
    
    \item \underline{Step 2d:} Evolve $N_l$ perturbed simulations and the baseline simulation simultaneously until $t = \tau_2$ to obtain solutions $\phi_i^{\tau_2}$ and $\phi_0^{\tau_2}$; Orthogonalize the solutions (this is the second and last orthogonalization) and compute $\delta \phi_i^{\tau_2} = \phi_i^{\tau_2} -\phi_0^{\tau_2}, i = 1,\cdots,N_l $
     \item \underline{Step 2e:} Compute the local LEs  as
     \[
        \lambda_i^k = \frac{1}{\tau_2-\tau_1} \text{log}(\frac{ \| \delta \phi_i^{\tau_2} \| }{ \| \delta \phi_i^{\tau_1} \|}), i = 1,\cdots,N_l
    \]
    \item \underline{Step 3:} Compute average LEs as $\lambda_i = \frac{1}{k}\sum_j^k \lambda_i^j$. Repeat step 2 using the orthogonalized solutions as initial perturbations until the LEs reach a converged estimate \cite{uncertaintyDNS-oliver}. Extract $\lambda_i$ and $\delta \phi_i , i = 1,\cdots,N_l$ as the LEs and LVs.
\end{itemize}

The choice of different parameters is dependent on the problem, and has been optimized using considerations mentioned in prior work on other related flows \cite{maliknumericsLE}. The initial time used to remove the effect of initialization, $\tau_1$ is set to $2 \times 10^{-5}$ s, while the time for which the perturbations are evolved, $\tau_2$, is set to $2 \times 10^{-4}$ s. Here, $\tau_2$ should be small enough to observe exponential growth of the perturbation and large enough to mitigate the computational cost of orthogonalization. These values were chosen using a sensitivity analysis. The same $\tau_1$ and $\tau_2$ were used for the coarse grid simulation. Note that the residence time based on the centerline velocity and the length of the domain is $1.1 \times 10^{-2}$ s for the D flame and $7.7 \times 10^{-3}$ s for the E flame. It is not necessary to use averaging times $\tau_2$ of the order of the flow through time as long as the number of algorithm cycles $k$ is large enough to approximate a long time average. Hence, the averaging times used here are much shorter than the flow-through times. The computations are repeated for $k=425$ for flame D, and $k=318$ for flame E, until the LEs are considered stationary. Since the LEs represent the average expansion rate, a sufficient number of samples along the traversed trajectories is necessary in order to obtain such average metrics with reasonable accuracy. In the results to be presented below, the uncertainty on the estimator of the mean value of the LEs will also be reported.


The Gram-Schmidt procedure leading to the orthonormal vectors is performed on matrices of size $N \times N_l$, which in this case will contain roughly 2.2 $\times 10^9$ elements for the Sandia D flame. Consequently, this algorithm requires extensive parallelization, and highly optimized implementation to ensure minimal transfer of data across computing nodes. The simulations were carried out on 2408 cores using MPI-based domain and matrix decomposition. The perturbations are recorded using the PETSc framework \cite{balay2016petsc} and the Gram-Schmidt procedure is implemented through the SLEPc module \cite{slepc}.

\section{Results and Discussion}
\label{sec:results}

\subsection{Characterization of Dimension of Attractor}
\label{sec:quantifyChaos}

As the Reynolds number is increased, the range of scales associated with a turbulent flow increases. Hence, in the pure dynamical system point of view, the number of total dimensions will increase. However, it is not guaranteed that the dimension of the attractor will follow the trend, since thermodynamics could impose strict relations that might constrain the chaoticity of the flow. Intuitively, the Sandia E flame can occupy more states in phase space due to the presence of extinction events. For instance, consider the conditional plots shown in Fig.~\ref{fig:sandiaIllustration}, which are projections of the entire phase-space into a two-dimensional sub-space. The presence of local extinction events lead to scatter near the stoichiometric mixture fraction with suppressed values for progress variable in the flame E simulations. Hence, the dimension of the attractor, which is the effective number of dimensions that dictate the evolution of the turbulent flame, is a crucial metric. Here, the Lyapunov spectrum of exponents is computed using the algorithm described in the previous section. The LEs are obtained over a total simulation time of $0.2$s for the the D flame and $0.12$s for the E flame. The coarse grid results are averaged over twice this amount of time. In the end, $300$ exponents were computed for both the D and the E flame.

\begin{figure}[h]
\center
\includegraphics[width=0.40\textwidth,trim={0cm 0cm 0cm 0cm},clip]{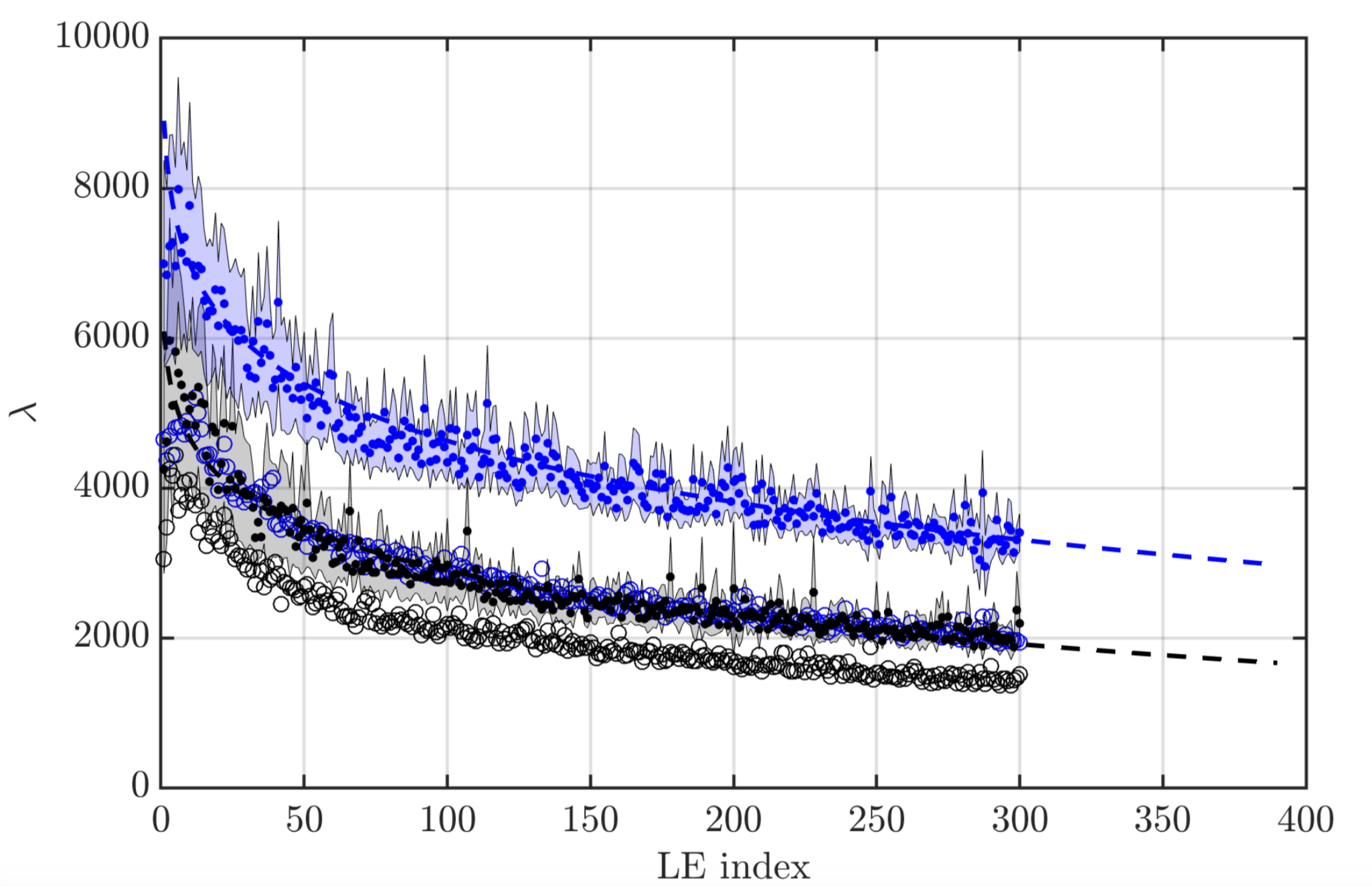}
\caption{Lyapunov exponents of the D flame (black symbols) and the E flame (blue symbols). Filled symbols are obtained with resolution $256 \times 128 \times 32$, empty symbols are obtained with resolution $172 \times 90 \times 32$. The shaded region denotes the uncertainty associated with the sampling errors for the exponents of the D flame (black shade) and the E flame (blue shade). The dashed lines (black for D flame and blue for E flame) are the inverse power law fit used to determine the attractor dimension.}
\label{fig:spectrumDE}
\end{figure}

Figure~\ref{fig:spectrumDE} shows the spectrum for both flames. For each exponent, an error estimate is defined by considering the time series of LEs and applying the central limit theorem while compensating for the lack of independence of the samples \cite{uncertaintyDNS-oliver}. The spectra exhibit interesting trends. First, there is a clear decay in the values with increase in the LE index, which indicates that there will exist LEs that will possess negative exponents, albeit at much higher LE indices. The trend is similar for both flames, indicating that the contribution of LE to the flow chaoticity decreases as the index of the LE increases. This is consistent with the notion of the strange attractor and the possibility of a smaller number of effective dimensions compared to $N$. In order to determine the dimension, a power fit is used. The attractor dimension can be obtained using the Kaplan-Yorke definition as:
\begin{equation}
    D = i + \frac{\sum_{1}^i \lambda_j}{ | \lambda_{i+1} |},
\end{equation}
where $\lambda_j$ denote the $j^{th}$ LE, and $i$ is the last index such that $\sum_1^i \lambda_j \geq 0 $. The estimation yields an attractor dimension of roughly 5000 for flame D and roughly 10500 for flame E. This estimate is obtained using the filtered solution and is expected to be lower than the dimension that would be obtained with fully resolved simulations. As an illustration, the spectrum obtained with the coarser resolution and shown in Fig.~\ref{fig:spectrumDE} would clearly lead to lower estimates of the attractor dimension. It shows that the dimension obtained here can be considered as a lower bound of the dimension of the fully resolved attractor.

Second, the LEs for flame E are larger than the corresponding LEs for flame D. Since the inverse of LE is a time-scale, this indicates that perturbations expand faster for in the flame E than the flame D. Given the same perturbation, the growth in these perturbations will be larger for flame E compared to flame D. The largest strain rates for flame D and E are roughly $2 \times 10^5$ s$^{-1}$ and $2.5 \times 10^5$ s$^{-1}$, respectively. On the other hand, the large scale time-scales can be estimated as approximately $10^{-3}$ s, based on the jet diameter and inflow turbulence fluctuations. Clearly, the LEs are much lower than the peak strain values, indicating that perturbation dynamics are controlled by length and time scales that lie in the inertial range of the turbulence cascade. Further, the increase in peak strain rate between the flames is lower than the corresponding increase in maximum LE, which is consistent with the findings of Mohan et al. \cite{mohan2017scaling} for homogeneous isotropic turbulence. This suggests that the increase in chaoticity is controlled by the turbulence properties rather than the chemical time scales. 

Finally, it is important to note that the larger LEs tend to have higher scatter in values, indicating that the local expansion rates in phase space change significantly. However, this scatter is reduced for the lower valued LEs (especially for flame D), which is an indication that there are stable sub-spaces where the perturbation expansion is dictated by flow properties that are location independent.

\subsection{Analysis of Lyapunov Vectors}

\begin{figure}[h]
\center
\includegraphics[width=0.49\textwidth,trim={0cm 0cm 0cm 0cm},clip]{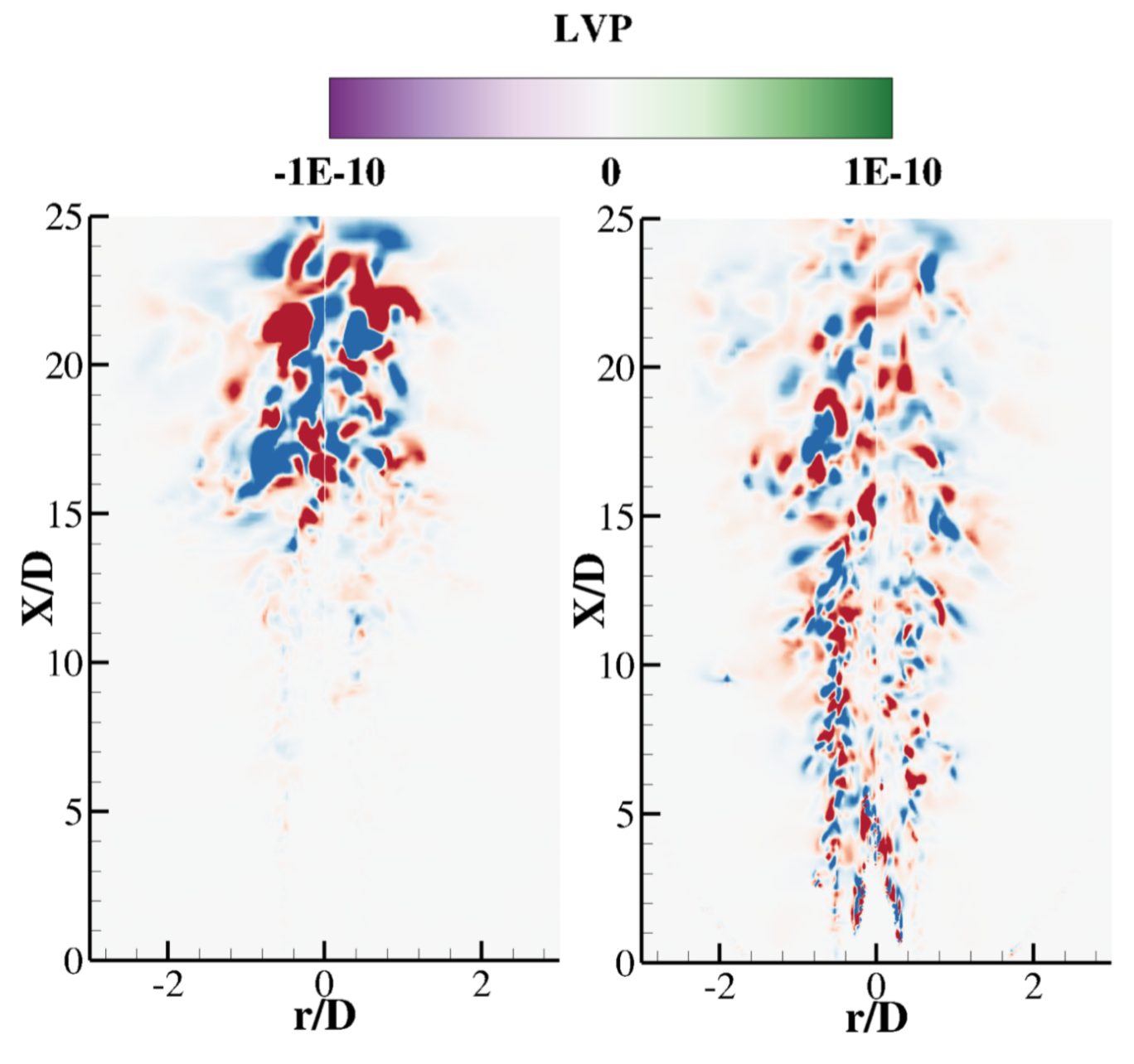}
\caption{Contours of the progress variable component of the first (left) and 300$^{th}$ LV (right) of the D flame.}
\label{fig:lv}
\end{figure}

The second component of the Lyapunov computations is the set of Lyapunov vectors, which are $N_l$ fields of the corresponding primary variables $N_s$. Each vector is orthogonal to every other vector by definition. Figure~\ref{fig:lv} shows the progress variable component of the first and 300$^{th}$ LVs of the D flame (in the following, this quantity is called LVP). Note that the magnitude range ($\pm 1 \times 10^{-10}$) is a function of the number of variables in the LV since the vector is normalized. While both fields have similar magnitudes, there is clear difference in their spatial distribution. The first LVP is clustered at particular locations of the jet. Further, this LV is highly unsteady (not shown here) and changes locations with time. The 300$^{th}$ LVP is spread along the shear layer that separates the fuel jet and the pilot. This suggests that the first LVP is controlled by the intermittent phenomena related to the breakdown of the turbulent jet core. The LVPs have a structure similar to strain/dissipation rate, with alternating negative and positive values. Further, the length scale of the structures decrease from the first to the 300$^{th}$ vector. Other LVs exhibit similar trends and are not shown here.

\begin{figure}[h]
\center
\includegraphics[width=0.40\textwidth,trim={0cm 0cm 0cm 0cm},clip]{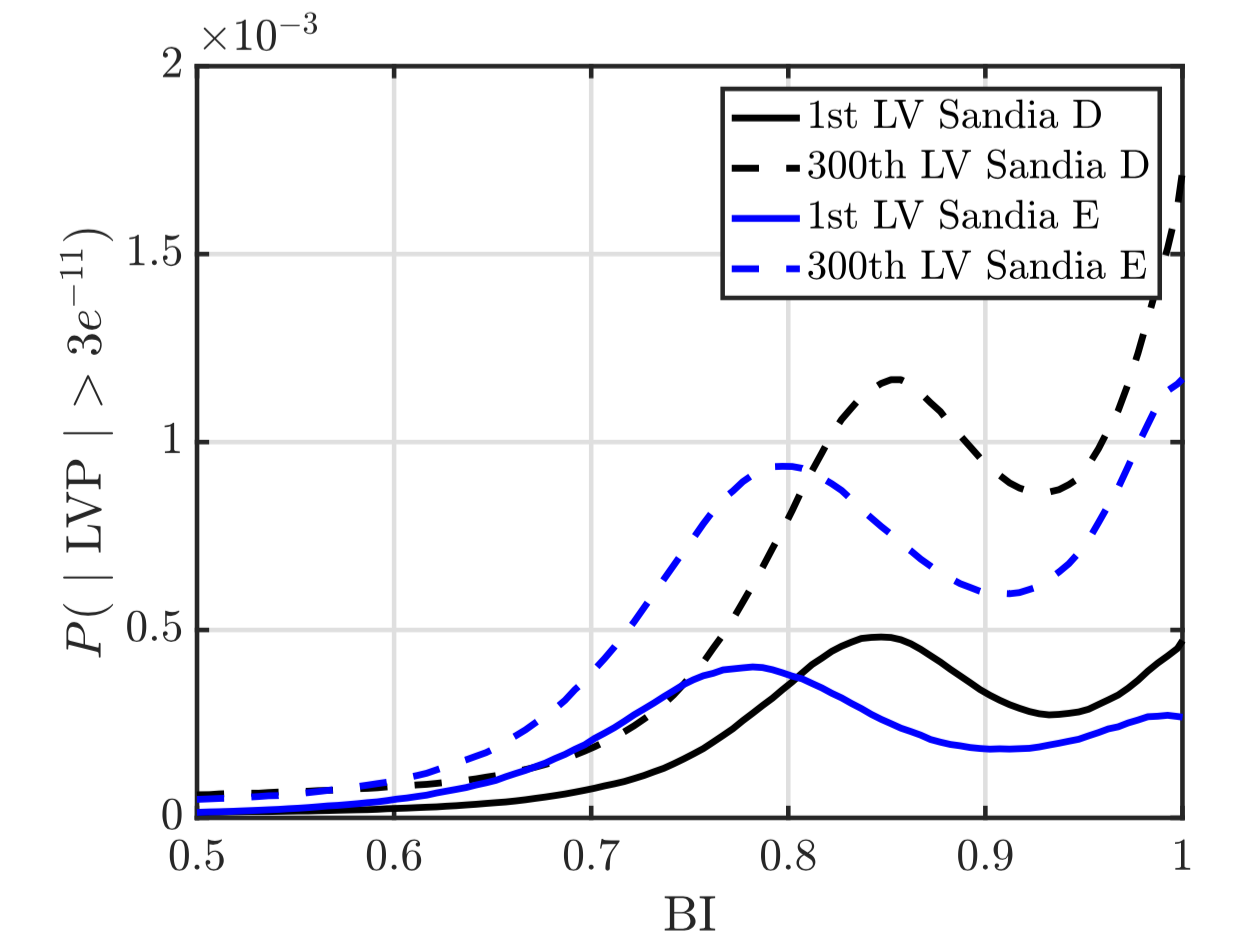}
\caption{Probability of encountering $|$LVP$|>3 \times 10^{-11}$ conditioned on the burning index. Solid lines denote the first LV, dashed lines denote the 300$^{th}$ LV. Black lines denote the Sandia D (obtained over 250 realizations) and blue lines denote the Sandia E (obtained over 160 realizations).}
\label{fig:pdfbi}
\end{figure}

Since the LVs are mutually orthogonal, it is expected that they identify different instability modes of the system. For the Sandia flame series, the main sources of chaoticity are extinction/re-ignition and turbulence. In  order to determine the relative roles of these physical mechanisms, the extinction process is first defined in terms of a burning index, a quantity that has been modified from the original definition \cite{xu2000pdf} to represent a field:
\begin{equation}
    BI(\boldsymbol{x}) = \frac{C(\boldsymbol{x})}{C_b(\boldsymbol{x})_{|Z_{mix}(\boldsymbol{x})}}, 
\end{equation}

where $BI(\boldsymbol{x})$ is the burning index field, $C(\boldsymbol{x})$ is the progress variable field and $C_{b_{|Z_{mix}}}$ is the progress variable at the local mixture fraction for a fully burning flamelet. A burning index close to 0 indicates extinction, while 1 denotes a fully burning solution. The maximum progress variable is obtained from the flamelets generated with the smallest strain rates. Figure.~\ref{fig:pdfbi} shows the probability of finding the absolute values of LVP components above a particular threshold conditioned on the burning index. For both flames, two different peaks exist: a peak at $BI = 0.85$ (flame D) or $0.8$ (flame E), and a peak at $BI=1$. The first peak is related to partially burning regions of the flame, and denotes a direction in phase-space along which any additional perturbation can cause flame to transition to an extinguished or re-ignited state. The second peak is related to turbulence induced jet motion in the fully burning part of the flame. The relative height of the peaks indicates which mode dominate each LV. It appears that the progress variable instabilities captured by the first LV are more related to extinction/reignition effects, while the instabilities captured by the last LV are more related to jet oscillation, i.e. turbulence induced instabilities. Note that the value of the threshold has little influence on the results, as long as infinitesimal values of LVP are filtered. Similar results were found with the coarser grid simulation (not included here) suggesting that these findings are independent of the spatial resolution used.

\begin{figure}[h]
\center
\includegraphics[width=0.40\textwidth,trim={0cm 0cm 0cm 0cm},clip]{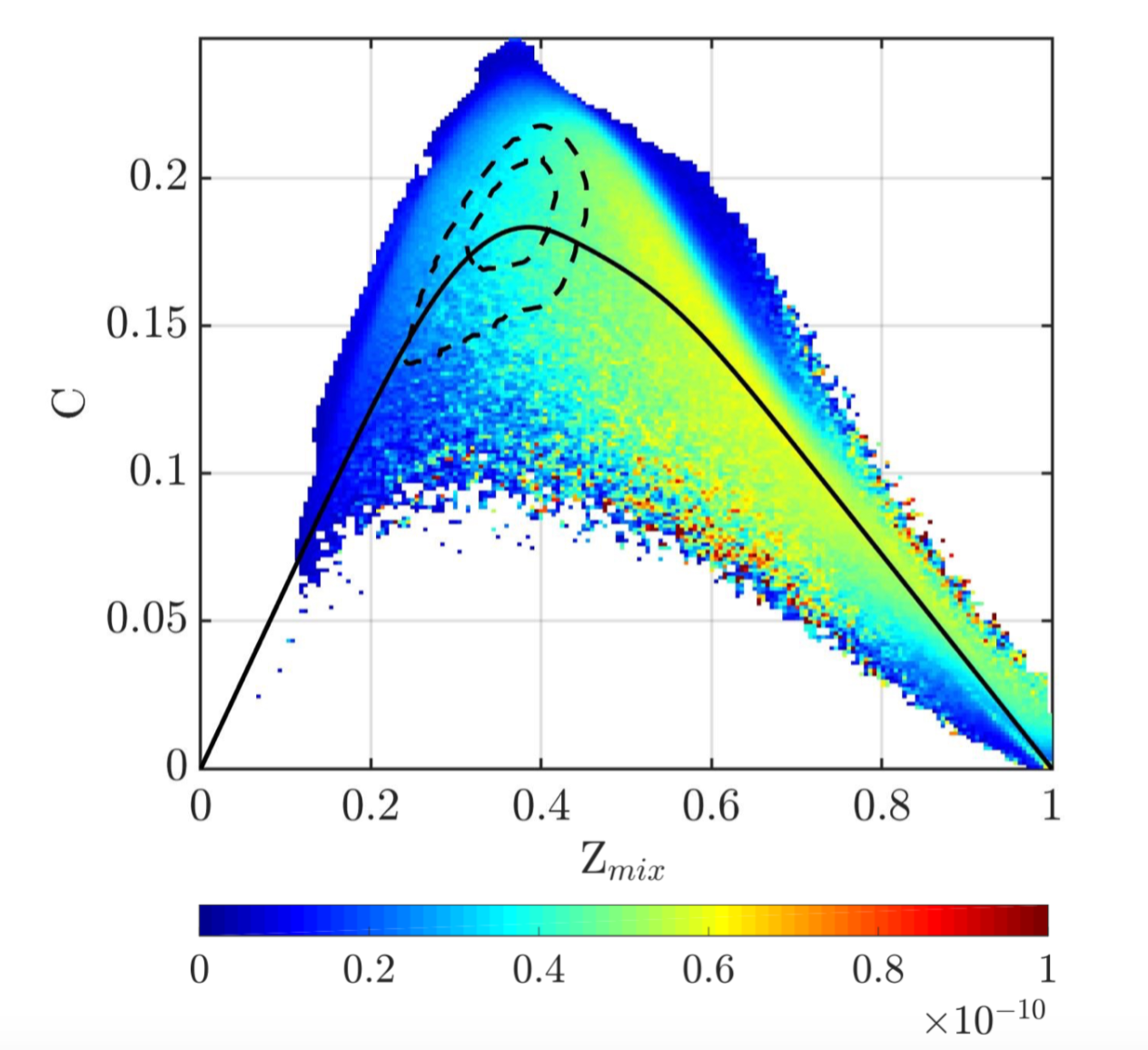}
\caption{Average conditioned on the mixture fraction and progress variables for $|$LVP$|$ for the 300$^{th}$ LV of the E flame. Average is taken over 160 realizations. Overlaid with $80\%$ of the fully burning flamelet of strain rate $a=87s^{-1}$ (solid line). Overlaid with 2 iso-contours of progress variable source term (dashed lines).}
\label{fig:meanZC}
\end{figure}

Figure~\ref{fig:meanZC} shows the distribution of the mean of the absolute value of LVP in the $(Z_{mix},C)$ space. The plot shows results consistent with the BI analysis (Fig.~\ref{fig:pdfbi}) as a peak in the LVP is visible for 0.8 of the fully burning flamelet (highlighted in the figure). This implies that the flame is the most responsive to perturbations in the region of phase space between fully burning and extinguished. The phase space plot also reveals that most of the perturbations are effective at a mixture fraction of around $Z_{mix}=0.6$, which corresponds to the shear layer between the inner jet and the pilot jet for $x/d<25$. It is also seen that the perturbations do not increase as much where the reaction is most active. For this flame, the reaction acts as a constraint on the chaoticity.

\section{Conclusions}
The focus of this study is to use a dynamical systems approach to understand perturbation dynamics of turbulent flames. This approach allows a full characterization of the chaotic dynamics associated with turbulent flows, which have broad ramifications for both modeling and control of such complex flows. The algorithms developed here can be directly applied to any practical flow, by replacing the computational solver used in the ensemble procedure. The ensemble-LES approach is used to estimate the dimension of the chaotic attractor for Sandia flames D and E. The results indicate the the dimension is at least 5000 (flame D) and 10500 (flame E) which are smaller than the full dimension of the state-space $N = 7.34 \times 10^6$. This is a significant result, which demonstrates that the strong thermodynamic relations imposed by the combustion process combined with heat-release-related-decrease in viscosity severely constrains the dynamics of the flow. 

In addition to the Lyapunov exponents, the LVs provide the directions along which perturbations grow the fastest in state-space. These modes are dictated by the interplay between extinction/re-ignition and the turbulence-induced jet motion. In particular, the strongest growth is related to the jet breakdown process, while other LVs are aligned with the extinction/re-ignition regions in the shear layer. Further, the perturbation growth modes exhibit location-dependent behavior in $(Z_{mix},C)$ space. The regions of extinction/re-ignition are shown to be highly chaotic, with perturbations being able to lead to full ignition or complete extinction with nearly equal probability. 

In conclusion, the Lyapunov theory presented here is a new tool for understanding the dynamics of turbulent flames. Although not discussed here, it has applications in prediction of macroscopic instabilities, as well as the prediction of rare events and trajectories in phase-space \cite{malikmcs10}. Further, it can provide a unified framework for understanding flame response to perturbations.


\section*{Acknowledgments}
\label{Acknowledgments}

This work was financially supported by an AFOSR research grant (FA9550-15-1-0378) with Dr.\,Chiping Li as program manager. The authors thank NASA HECC for generous allocation of computing time on NASA Pleiades machine.

\bibliographystyle{elsarticle-num}



\end{document}